\documentclass[12pt,preprint]{aastex}

\newcommand{\fer}{{\it Fermi}}
\newcommand{\wse}{{\it WISE}}

\newcommand{\gstrp}{{\it WGS}}
\usepackage{graphicx}
\usepackage{longtable}

\slugcomment{version \today: fm}

\shorttitle{Unidentified gamma-ray sources: hunting $\gamma$-ray blazars}
\shortauthors{F. Massaro, R. D'Abrusco, G. Tosti, M. Ajello, Paggi, A., D. Gasparrini}

\begin{document}
\title{Unidentified gamma-ray sources: hunting $\gamma$-ray blazars}
\author{F. Massaro\altaffilmark{1}, R. D'Abrusco\altaffilmark{2}, G. Tosti\altaffilmark{3,4}, M. Ajello\altaffilmark{1}, A. Paggi\altaffilmark{2}, D. Gasparrini\altaffilmark{5}.}

\altaffiltext{1}{SLAC National Laboratory and Kavli Institute for Particle Astrophysics and Cosmology, 2575 Sand Hill Road, Menlo Park, CA 94025}
\altaffiltext{2}{Harvard - Smithsonian Astrophysical Observatory, 60 Garden Street, Cambridge, MA 02138}
\altaffiltext{3}{Dipartimento di Fisica, Universit\`a degli Studi di Perugia, 06123 Perugia, Italy}
\altaffiltext{4}{Istituto Nazionale di Fisica Nucleare, Sezione di Perugia, 06123 Perugia, Italy}
\altaffiltext{5}{ASI Science Data Center, ESRIN, I-00044 Frascati, Italy}

\begin{abstract}
One of the main scientific objectives of the ongoing \fer\ mission is unveiling the nature of the unidentified $\gamma$-ray sources (UGSs).
Despite the large improvements of \fer\ in the localization of $\gamma$-ray sources with respect to the past $\gamma$-ray missions, 
about one third of the \fer-detected objects are still not associated to low energy counterparts. 
Recently, using the Wide-field Infrared Survey Explorer (\wse) survey, we discovered that blazars, the rarest class of Active Galactic Nuclei 
and the largest population of $\gamma$-ray sources, 
can be recognized and separated from other extragalactic sources on the basis of their infrared (IR) colors.
Based on this result, we designed an association method for the $\gamma$-ray sources 
to recognize if there is a blazar candidate within the positional uncertainty region of a generic $\gamma$-ray source.
With this new IR diagnostic tool, we searched for $\gamma$-ray blazar candidates associated to the UGS sample of the 
second \fer\ $\gamma$-ray catalog (2FGL). 
We found that our method associates at least one $\gamma$-ray blazar candidate as a counterpart each of 156 out of 313 UGSs analyzed.
These new low-energy candidates have the same IR properties as the blazars associated to $\gamma$-ray sources in the 2FGL catalog.
\end{abstract}

\keywords{galaxies: active - galaxies: BL Lacertae objects -  radiation mechanisms: non-thermal}

\section{Introduction}
\label{sec:intro}
More than half of the $\gamma$-ray sources detected by {\it Compton} Gamma-Ray Observatory (CGRO),
and present in the third EGRET (3EG) catalog were not 
associated with known counterparts seen at low energies \citep{hartman99}.
Whatever the nature of the unidentified $\gamma$-ray sources (UGSs), 
these objects could provide a significant contribution to the isotropic gamma-ray background (IGRB) \citep[e.g., ][]{abdo10a}.
Solving the puzzle of the origin of the UGSs, together with a better knowledge of other 
IGRB contributions estimated from known sources, 
is crucial also to constrain exotic high-energy physics phenomena, 
such as dark matter signatures, or new classes of sources.

With the advent of the \fer\ mission the localization of $\gamma$-ray sources 
has significantly improved with respect to the past $\gamma$-ray missions, thus 
simplifying the task of finding statistically probably counterparts at lower energies.
New association methods also have been developed and applied, so that the number of 
UGSs has significantly decreased with respect to the 3EG catalog  \citep{hartman99};
however, according to the second \fer\ $\gamma$-ray catalog (2FGL), about one third 
of detected gamma-ray sources in the energy range above 100 MeV is still unassociated \citep{abdo11}.  
It is worth noting that the most commonly detected sources in the $\gamma$-ray sky, since the epoch of CGRO, are blazars,
one of the most enigmatic classes of Active Galactic Nuclei (AGNs) \citep[e.g.,][]{hartman99}.
Within the 2FGL, there are 576 UGSs out of a total number of 1873 sources detected, 
while among the 1297 associated sources, $\sim$ 1000 have been associated with AGNs \citep{abdo11,ackermann11a}.

Blazar emission extends over the whole electromagnetic spectrum and is generally interpreted as non-thermal radiation
arising from particles accelerated in relativistic jets closely aligned to the line of sight \citep{blandford78}.
They come in two flavors: the BL Lac objects, with featureless optical
spectra or only with absorption lines of galactic origin and weak and narrower emission lines, 
and the Flat Spectrum Radio Quasars, with a optical spectra showing broad emission lines.
In the following, we indicate the former as BZBs and the latter as BZQs, respectively, 
according to the ROMA-BZCAT\footnote{http://www.asdc.asi.it/bzcat/} nomenclature \citep{massaro09,massaro10,massaro11a}.

The first step to improve our knowledge on the origin of the UGSs and 
of their associations with low-energy counterparts,
is to recognize those that could have a blazar within their $\gamma$-ray positional uncertainty regions.

Recently, we developed a procedure to identify 
blazars using their infrared (IR) colors within the preliminary data release 
of the Wide-field Infrared Survey Explorer (\wse) survey \citep{wright10}
\footnote{http://wise2.ipac.caltech.edu/docs/release/prelim/}.
In particular, we discovered that the IR color space distribution of
the extragalactic sources dominated 
by non-thermal emission, as blazars, can be used to distinguish such sources
from other classes of galaxies and/or AGNs and/or galactic sources \citep[][hereinafter Paper I]{massaro11b}.
We also found that $\gamma$-ray emitting blazar delineate a narrow, 
distinct region of the IR color-color plots, denominated as the 
\wse\ Gamma-ray blazar Strip (\gstrp) \citep[][hereinafter Paper II]{dabrusco12}.
There is a peculiar correspondence 
between the IR and $\gamma$-ray spectral properties of the blazars detected in the 2FGL (Paper II). 
Then, on the basis of our previous investigation of these IR-$\gamma$-ray properties of blazars,
we built a parametrization of the \gstrp\ to evaluate how many AGNs of Uncertain type (AGUs)
have a counterpart associated with a $\gamma$-ray blazar candidate in the 
2FGL \citep[][hereinafter Paper III]{massaro12a}.

In this paper, we present a new association method based on the IR colors
of the $\gamma$-ray emitting blazars and the \gstrp\ parametrization.
Then we apply this new association procedure to search for $\gamma$-ray blazar candidates 
within the $\gamma$-ray positional error regions of the UGSs.
One of the main advantages of our method is that it reduces 
the number of potential counterparts for the UGSs and provides their positions with arcsec resolution, thus 
restricting the search regions for future followup observations necessary to confirm 
their blazar nature. Unfortunately, only a restricted number of UGSs falls within the 
portion of the sky currently covered by the IR observations of the \wse\ Preliminary Data Release 
corresponding to $\sim$ 57\% of the whole sky.
Then, when the \wse\ survey will be completely released in March 2012\footnote{http://wise2.ipac.caltech.edu/docs/release/allsky/},
it will be possible to apply the method to the whole sky,
even in regions not covered at radio, optical and X-ray frequencies,
where the other methods for establishing counterpart associations for the 2FGL cannot be used. 

This paper is organized as follows: in Section~\ref{sec:sample} we describe the 
samples used in our investigation; in Section~\ref{sec:method} we illustrate 
the new association method; then, in Section~\ref{sec:ugs} we apply the new association technique
to the UGSs and describe the subset of sources that has been associated with $\gamma$-ray blazar candidates. 
In Section~\ref{sec:comparison}, we also compare our results with those found adopting different statistical approaches
for a subsample of UGSs. Finally, conclusions are presented in Section~\ref{sec:summary}.

\section{The sample selection}
\label{sec:sample}
To build our association procedure we considered a sample of blazars selected from the combination of the 
ROMA-BZCAT \citep{massaro09,massaro10} and the 2FGL \citep{abdo11},
as described and used in Paper II and used in Paper III to parametrize the \gstrp, denoted the 2FB sample.
It contains 284 $\gamma$-ray blazars (135 BZBs and 149 BZQs) 
that have optical and radio counterparts as reported in the ROMA-BZCAT,
and also having a \wse\ counterpart within 2.4 $^{\prime\prime}$ radius (see Paper I and III).
The blazars in the 2FB sample are detected by \wse\ with a signal to noise ratio higher than 7 in at least one band
and do not have any upper limits in all the \wse\ bands.
We excluded from our analysis all the blazars with a \fer\ analysis flag, 
according to the 2FGL and the 2LAC \citep{abdo11,ackermann11a}.
The blazars of uncertain type (BZUs)
have been excluded from our analysis, while the BL Lac candidates have been considered as BZBs.
More details on the 2FB sample and the source selections are given in Papers II and III.
 
Then, we applied our association procedure to the sample of the UGS defined as follows.
The number of UGSs in the 2FGL is 576, but only 410 of these $\gamma$-ray sources
lie in the region of the sky available in the \wse\ Preliminary Data Release.
These sources can be analyzed according to our method based on the IR \wse\ colors.
We adopted a more conservative selection
restricting our sample to 313 UGSs out of 410, excluding sources with a \fer\ analysis flag,
since these sources might not be real and/or could be affected by analysis artifacts \citep[see e.g.][for more details]{abdo11}.

\section{The \gstrp\ association method}
\label{sec:method}
In Paper III, working on the AGUs, we built the \gstrp\ parametrization 
to verify if the low-energy counterparts of the AGUs, associated in 2FGL, 
is consistent with the \gstrp, so being a $\gamma$-ray blazar candidate.
With respect to the previous analysis, the following proposed association procedure  
aims at providing new $\gamma$-ray blazar candidates, possible counterparts of the UGSs, 
that lie within their $\gamma$-ray positional uncertainty regions,
on the basis of our previous results on the IR-$\gamma$-ray blazar properties.
In this Section, we report the basic details of our \gstrp\ parametrization together with the
definition of different classes of $\gamma$-ray blazar candidates.
Then we describe our new association procedure.

\subsection{The \gstrp\ parametrization}
\label{sec:parameter}
In Paper II, we found that $\gamma$-ray emitting blazars (i.e., those in the 2FB sample)
cover a narrow region in the 3D color space built with the \wse\ magnitudes
delineating the so-called \wse\ Gamma-ray blazar Strip (\gstrp).
 
In Paper III, using the 2FB sample, we presented the parametrization of the \gstrp\ 
based on the {\it strip parameter} $s$. This parameter, ranging between 0 and 1, provides a measure of the
distance between the \gstrp\ and the location of a \wse\ source in the three dimensional IR color parameter space.
For example, sources with high values of $s$ (e.g., $\geq$ 0.50) are consistent with the \gstrp.
We also distinguished between \wse\ sources that lie in the subregion of the \gstrp\
occupied by the BZBs and BZQs using the $s_b$ and $s_q$ parameters separately (Paper III).

The IR color space has been built using the archival data available in the 2011 \wse\ Preliminary Data Release, 
in four different bands centered at 3.4, 4.6, 12, and 22 $\mu$m 
with an angular resolution of 6.1, 6.4, 6.5 \& 12.0$^{\prime\prime}$, respectively and achieving 5$\sigma$ 
point source sensitivities of 0.08, 0.11, 1 and 6 mJy. 
In addition, the absolute (radial) differences between \wse\ source-peaks and ``true" astrometric positions 
anywhere on the sky are no larger than $\sim$ 0.50, 0.26, 0.26, and 1.4$^{\prime\prime}$ in the
four \wse\ bands, respectively \citep{cutri11}\footnote{http://wise2.ipac.caltech.edu/docs/release/prelim/expsup/sec2\_3g.html}.

\subsection{$\gamma$-ray blazar candidate definition}
\label{sec:association}
Based on the $s_b$ and $s_q$ distributions of all \wse\ sources in different random regions of the sky,
at both high and low Galactic latitudes (Paper III), 
the critical threshold of the $s$ parameters, used 
to define the above classes, have been arbitrarily determined on the basis of the following considerations:
\begin{itemize}
\item{class A: \wse\ sources with 0.24 $<s_b<$ 1.00 and 0.38 $<s_q<$ 1.00;}
\item{class B: \wse\ sources with 0.24 $<s_b<$ 1.00 or 0.38 $<s_q<$ 1.00;}
\item{class C: \wse\ sources with 0.10 $<s_b<$ 0.24 and 0.14 $<s_q<$ 0.38.}
\end{itemize}
All the \wse\ sources with $s_b<$0.10 or $s_q<$0.14 are considered {\it outliers} of the \gstrp\ and, 
for this reason, discarded. 
All the above thresholds are then used to select the \wse\ sources that are associated to the UGSs and that can 
be considered potential $\gamma$-ray blazar candidates.

The above choice of threshold have been adopted for the analysis of the $\gamma$-ray blazar 
content within the AGUs (Paper III).
From the distributions of the $s_b$ and $s_q$ parameters for the generic IR \wse\ sources,
we note that 99.9\% of them have $s_b<$0.24 and $s_q<$0.38.
Then, for the BZBs in the 2FB sample only 6 sources out of 135 have $s_b<$ 0.24, and in the case of the BZQs
only 33 sources out of 149 show $s_q$ values lower than 0.38.
We also note that 99.0\% of the generic IR \wse\ sources have $s_b<$0.10 and only 2 BZBs are below this value,
while 97.2\% of the generic IR \wse\ sources together with only 5 BZQs out of 149 have $s_q<$0.14.

The \wse\ objects of class A are the most probable blazar counterpart of the unidentified $\gamma$-ray 
sources, because their WISE colors are more consistent with the \gstrp\ 
in both the BZBs and BZQs subregions than the colors of sources of class B or C.
Based on the distributions of the $s_b$ and $s_q$ parameters for \wse\ sources in 
random region of the sky, the sources of class A are, as expected, rarer than the sources belonging to
the other two classes (see Section~\ref{sec:ugs} for more details).

\subsection{The association procedure}
\label{sec:procedure}
\begin{figure}[]
\includegraphics[height=6.8cm,width=9.5cm,angle=0]{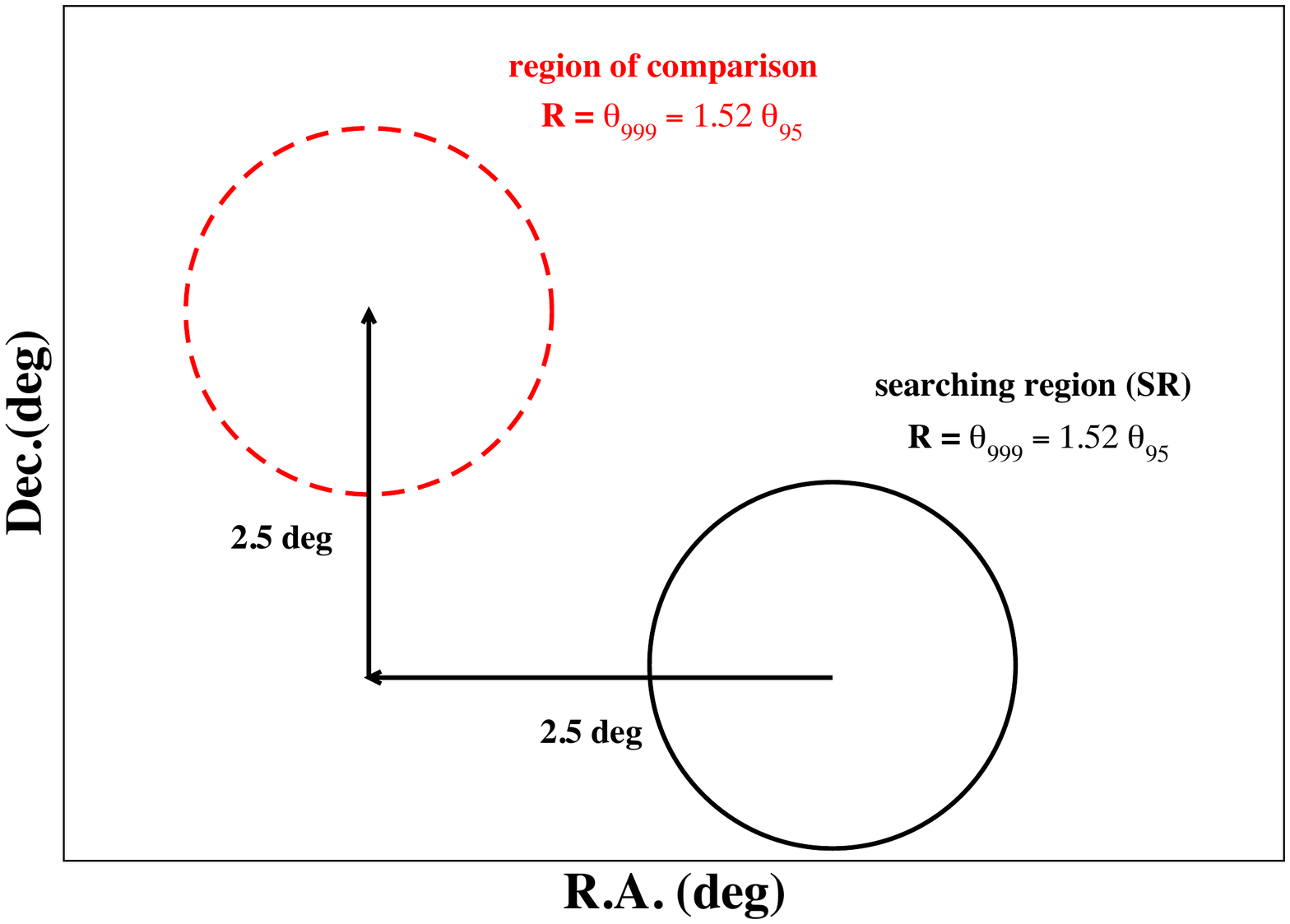}
\caption{The position of the region of comparison (ROC) for a generic \fer\ source,
with respect to the searching region (SR) centered on the position reported in the 2FGL catalog.
The radius of both regions is $R=\theta_{999}$ and they are separated by 2.5$\sqrt{2}$ deg of distance.}
\label{fig:roi}
\end{figure}
For each unidentified $\gamma$-ray source we defined the {\it searching region} (SR) corresponding 
to a circular region of radius $R$=$\theta_{999}$, 
centered on the position given in the 2FGL,
where $\theta_{999}$ is the major axis of the elliptical source location region 
corresponding to the 99.9\% level of confidence.
In addition, we also considered a {\it region of comparison} (ROC) 
defined as a circular region of the same radius $R$, 
but lying at 2.5$\sqrt{2}$ deg angular distance from the 2FGL position. 
A schematic view of the locations of the SR and the ROC is shown in Figure~\ref{fig:roi}.

Successively, for every unassociated gamma-ray source in the 2FGL catalog,
we ranked all the \wse\ sources within its SR 
on the basis of the classification described above and we selected as $\gamma$-ray blazar candidates 
the positionally closest sources with the highest class.
In our analysis we considered only sources of the \wse\ preliminary catalog detected in all the four \wse\ bands, without any upper limit.

The ROCs are used to assess the association confidence that a \wse\ source in a random region in the sky,
where no $\gamma$-ray source is located, has IR colors compatible with the WGS.
To provide an estimate of the association confidence, 
we considered the distribution of the strip parameters $s_b$ and $s_q$ for all the \wse\ sources within each ROC
associated to an UGS.
For these \wse\ sources we estimated the confidence $\pi$ that a generic \wse\ source belongs to the same class
as the $\gamma$-ray blazar candidate selected within the SR.
Thus the $\pi$ value will be expressed as the ratio between the number of \wse\ sources of a particular class
and the total number of \wse\ sources that lie in the ROC.

\subsection{Testing the association method with blazars}
\label{sec:test}
We performed a test to evaluate the completeness of our association method searching for the $\gamma$-ray blazar
candidates that are potential counterparts of the 2FB sample,
and verifying whether our procedure correctly finds the same associations as in the 2FB sample.

Assuming that the 284 blazars in the 2FB sample have been associated to the real low-energy counterparts, 
we run our association procedure considering the IR colors for all the \wse\ sources within the SRs for all these sources.
We found that for the population of BZBs, consisting of 135 BL Lacs,
our association procedure is able to recognize 123 sources as the 2FGL, 62 of class A, and 61 of class B.
Within the remaining 12 BZBs, 3 objects are associated to WISE sources of higher class than the original
2FGL associated sources, while for 9 sources we only found outliers of the \gstrp\ within their SRs.

For the BZQs, our method finds the same associations as in the
2FGL catalog for 124 of the sources, with 85 sources classified as class A, 32 classified as class B and 7 as class C. 
For the remaining 25 sources, we found 11 outliers and 14 $\gamma$-ray sources associated to  
a \wse\ source with higher classes. 

Our procedure re-associates 247 out of 284 $\gamma$-ray blazars of the 2FB sample in agreement with the 2FGL analysis, 
with a completeness of 87.0\% (91.0\% for the BZBs and 83.0\% for the BZQs). 
We found that 7.1\% are outliers of the \gstrp, but this number can be expected
because the \gstrp\ parametrization was built to require at least 90\% of the 2FB sources 
inside each 2-dimensional \gstrp\ projection (see Paper III for more details).

It is interesting to note that 17 out of 284 $\gamma$-ray sources in the 2FGL have a ``better", on the basis of our method, 
$\gamma$-ray blazar candidate within the SR.
These associations need to be verified with followup observations, as for example in the X-rays,
and a deeper analysis to check their reliability relative to the 2FGL association method will be performed in a forthcoming paper\citep{massaro12b}.

\section{Results}
\label{sec:ugs}
\begin{figure}[]
\includegraphics[height=6.0cm,width=9.7cm,angle=0]{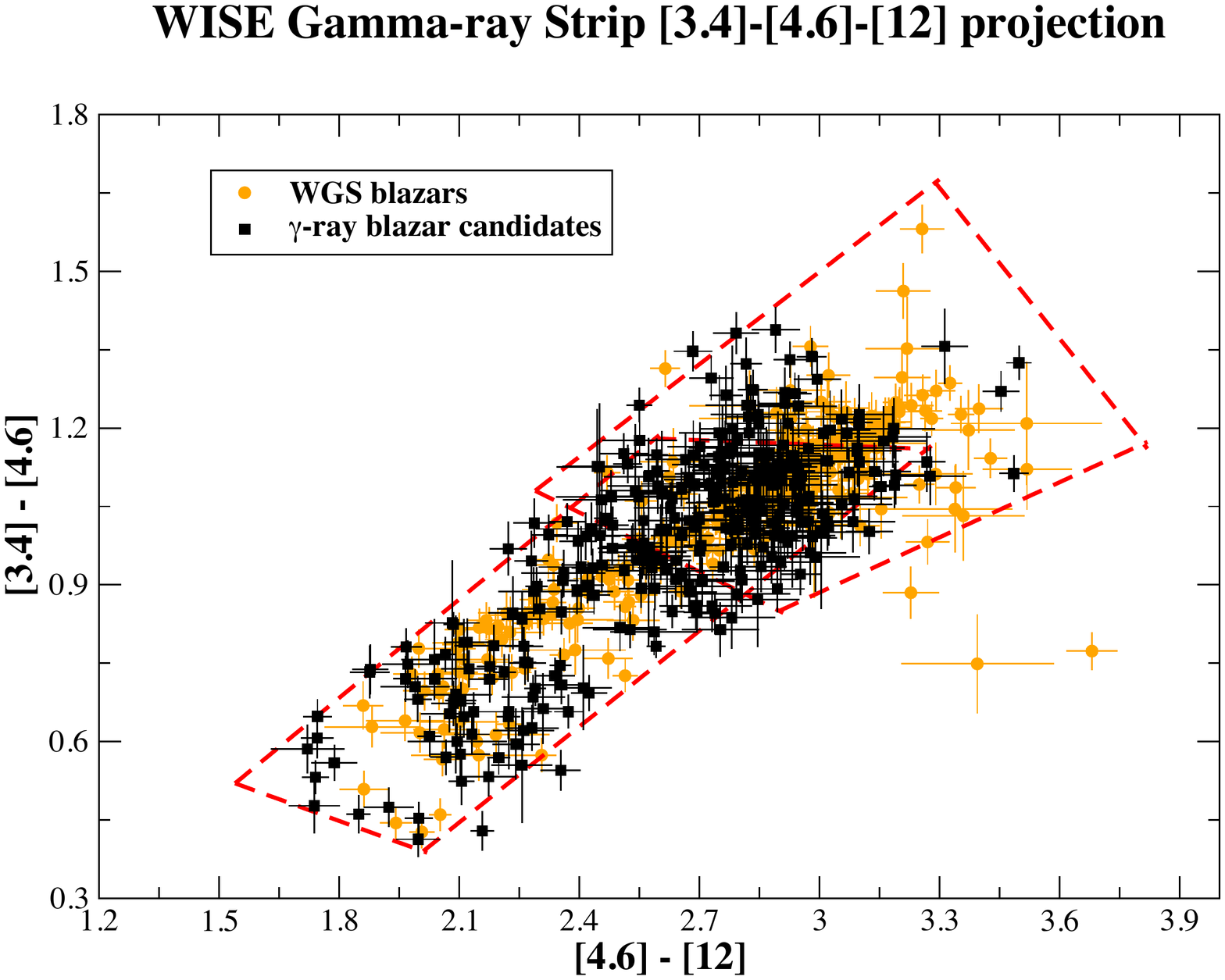}
\caption{The [3.4]-[4.6]-[12] $\mu$m 2D projection of the \gstrp\ is shown.
Red dashed lines show the boundaries of the \gstrp\ used in our analysis (see Paper III for more details).
The orange background filled circles are the blazars associated with the 2FGL constituting the 2FB sample
while the balck filled circles indicate the 156 $\gamma$-ray blazars that have been associated by our procedure.}
\label{fig:strip_pln1}
\end{figure}

The application of our association procedure to the 313 UGSs selected from the 
2FGL (see Section~\ref{sec:sample} for more details),
led to the associations of 156 UGSs with a low-energy candidate $\gamma$-ray blazar counterpart within their SRs.
According to our criteria (see Section~\ref{sec:association}), these 156 new associations consist of 
44 sources of class A, 74 of class B and 38 of class C. 
Thus our procedure finds associations with likely $\gamma$-ray blazar candidates for 49.8\% of the UGSs analyzed.
We also list of all the $\gamma$-ray blazar candidates with lower class for each UGSs,
if more than one is present within the SRs.
Among these 156 new associations, for 86 sources,
12 of class A, 43 of class B and 31 of class C, have only a single $\gamma$-ray blazar candidate within the SR.
In Figure~\ref{fig:strip_pln1} we show the \wse\ colors of the 156 $\gamma$-ray blazar candidates 
in comparison with those of the blazars in the 2FB sample for the [3.4]-[4.6]-[12] $\mu$m 2D projection of the \gstrp.

By restricting our sample of UGSs only to those at high Galactic latitudes, i.e. $|b|>$15$^\circ$,
we found a $\gamma$-ray blazar candidate for 72 UGSs, 16 of class A, 29 of class B and 27 of class C;
where for 34 out of these 74, the low energy counterpart associated with our method is univocal.
\begin{figure}[]
\includegraphics[height=6.0cm,width=9.7cm,angle=0]{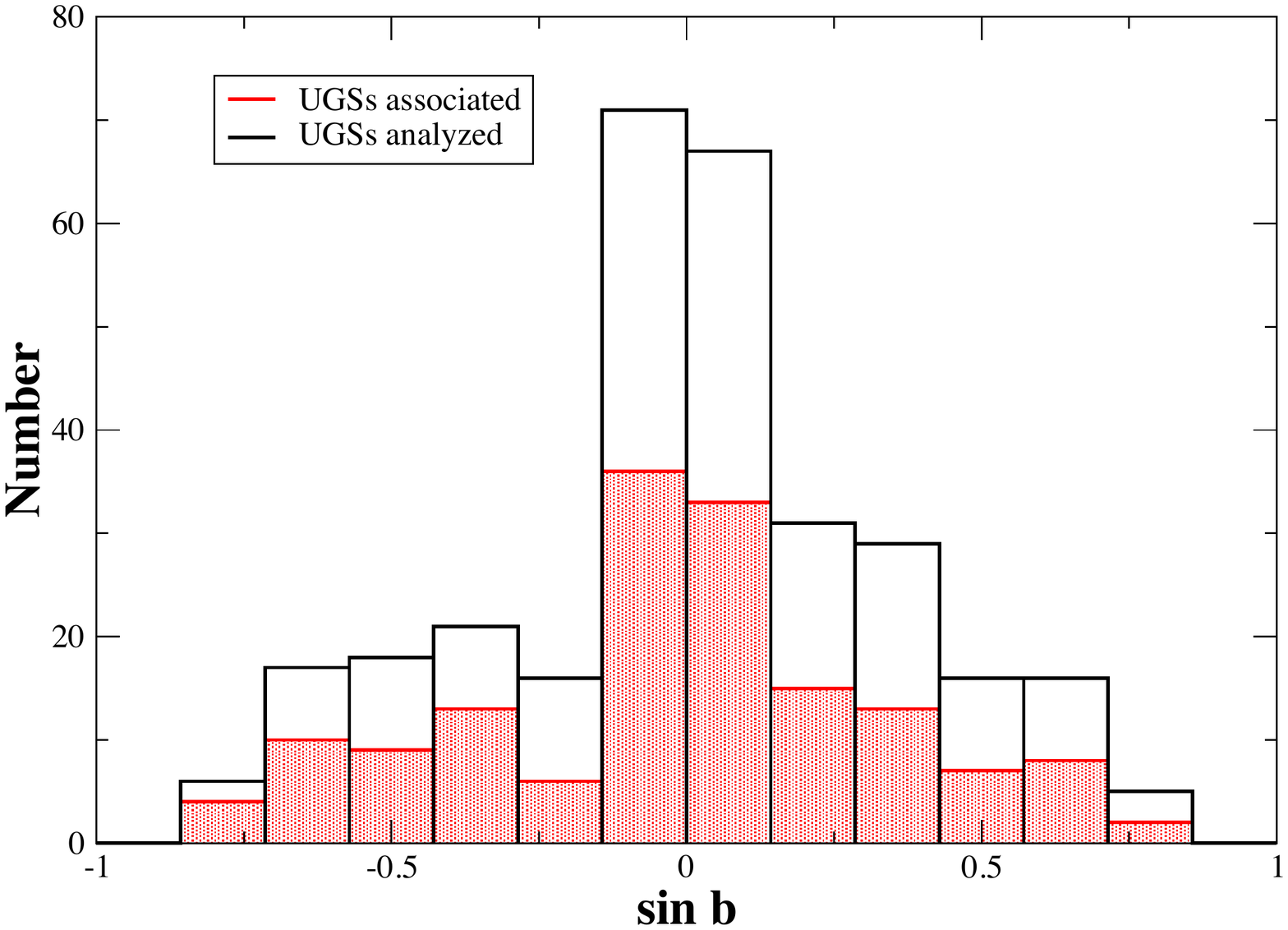}
\caption{The distribution of the Galactic latitude
for all the UGSs analyzed in comparison with that for the 156 associated by our procedure.}
\label{fig:glat_distrib}
\end{figure}
In Figure~\ref{fig:glat_distrib}, we shown the distribution of the Galactic latitude (i.e., sin $b$ )
for all the UGSs analyzed in comparison with those 156 associated by our method.
At high Galactic latitude, the method seems to be less efficient given the ratio between the number of UGSs analyzed and those associated.
This could be due to the non uniform exposure of the archival \wse\ observations in the 
\wse\ Preliminary Data Release\footnote{http://wise2.ipac.caltech.edu/docs/release/prelim/figures/prelim\_3x3-w1-equ.jpg}, 
and will be re-analyzed once the whole \wse\ archive will be available.
In addition, we note that our association method could be more efficient at low Galactic latitudes
where the blazar catalogs, as the ROMA-BZCAT, are less complete \citep{massaro09}.

\begin{figure}[!b]
\includegraphics[height=6.0cm,width=9.7cm,angle=0]{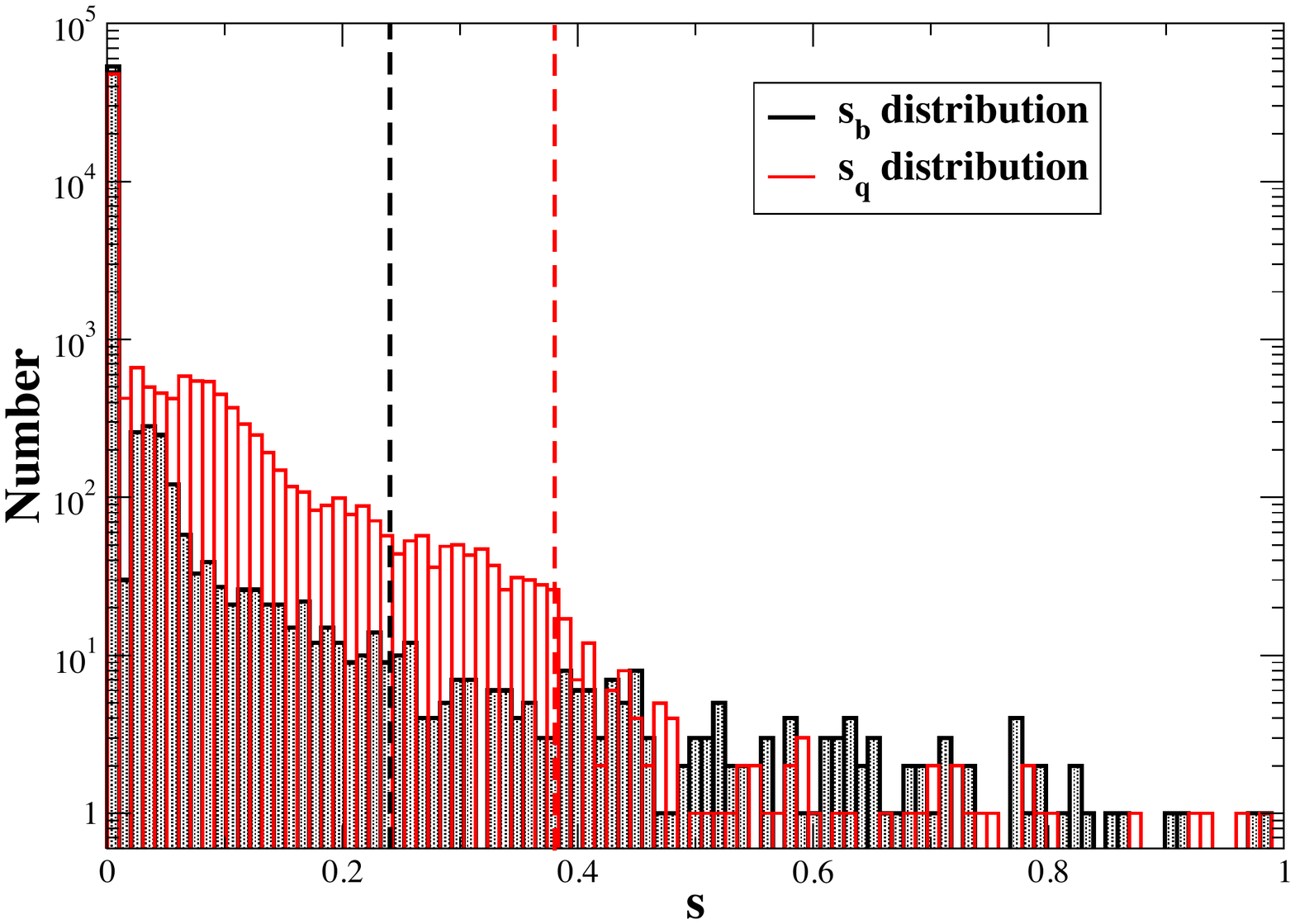}
\caption{The distribution of the $s_b$ (black) and $s_q$ (red) parameters
for all the 55195 \wse\ source within all the ROCs defined for the 313 UGSs analyzed.
The vertical lines corresponds to the thresholds for the $s_b$ and $s_q$ parameters 
to determine the blazar classes (see Section~\ref{sec:association}).}
\label{fig:histogram}
\end{figure}

We also remark that within the 313 regions of comparison chosen for the UGSs 
there are 55195 \wse\ sources, but only 49 of class A, 213 of class B and 129 of class C, all of them
detected in all four \wse\ bands and with a signal to noise ratio higher than seven in at least one band,
as the blazars in the 2FB sample.
The distributions of the $s_b$ and $s_q$ parameters for all the 55195 \wse\ sources within the 
313 ROCs are shown in Figure~\ref{fig:histogram}.
A blind search of all the possible $\gamma$-ray blazar candidates in the \wse\ archive on the basis of the \gstrp\ properties
will be performed once it will be completely available \citep{massaro12b}. 
However, the $s_b$ and $s_q$ distributions reported in Figure~\ref{fig:histogram}  
strongly suggest that the density of \wse\ blazar candidates is low over the sky.

In Table~\ref{tab:example} we show three cases of \wse\ 
sources that have been associated with our procedure to 
three UGSs. We report both the $s_b$ and $s_q$ values, the \gstrp\ class 
and the association confidence $\pi$.
In this example, the source 2FGL J0038.8+6259 is associated to one \wse\ source of a class 
A, J003818.70+630605.0, that has been selected as a single $\gamma$-ray blazar candidate out of 791 \wse\ sources 
within its SR.
The corresponding association confidence $\pi$, expressed in 
terms of number of sources with an higher $s_b$ or $s_q$ values 
than J003818.70+630605.0 within the region of comparison and estimated considering 830 \wse\ sources, is 2/830.

Similarly, the source 2FGL J0616.6+2425 has been associated to the \wse\ 
source J061609.79+241911.0, that belongs to class B
with a low association confidence estimated on 5465 \wse\ sources in the region of comparison. 
The 2FGL source 2FGL JJ0312.8+2013 has a \wse\ class C source associated 
following our procedure, with a lower confidence of
finding a similar source in a region of comparison where there are 512 \wse\ sources.

Within the 313 UGS analyzed there are 14 sources that have a variability index \citep{abdo11}
higher than the value of 41.6 corresponding to the 99\% of confidence that the source is variable.
It is worth noting that 13 out of these 14 variable UGSs have been successfully associated here
with a $\gamma$-ray blazar candidate, strongly supporting the blazar nature.

\begin{table}
\tiny\caption{An example of three UGS associations.}
\begin{tabular}{|lrlcccc|}
\hline
2FGL & Sources         & \wse\  & $s_b$ & $s_q$ & class & $\pi$ \\
name &  in SR & name &           &             &          \\
\hline
\noalign{\smallskip}			         
J0038.8+6259  &   791  & J003818.70+630605.2 & 0.89 & 0.99 & A & 2/830  \\
J0616.6+2425  &  6021 & J061623.95+241809.2 & ---    & 0.57  & B & 1/5465 \\ 
J0312.8+2013  &    453 & J031223.00+200749.5 & 0.19 & 0.15 & C & 1/512\\ 
\hline
\noalign{\smallskip}			           
\end{tabular}
\label{tab:example}
\end{table}

The entire list of the UGSs analyzed can be found in Table 2.
For each UGS, we report all the 
$\gamma$-ray blazar candidates with their IR colors (i.e., $c_{12}$ = [3.4]-[4.6] $\mu$m, 
$c_{23}$ = [4.6]-[12] $\mu$m and $c_{34}$ = [12]-[22] $\mu$m, 
together with their errors, $\sigma_{12}$, $\sigma_{23}$, $\sigma_{34}$, respectively), 
the distances in arc seconds between the $\gamma$-ray position and the selected \wse\ source, the $s_b$ and $s_q$ values,
the class and the association confidence $\pi$ that there is 
a \wse\ source of the same class within the ROC (see Section~\ref{sec:association}).

In addition, we found that there are 157 unidentified $\gamma$-ray sources that
do not have clear $\gamma$-ray blazar counterpart within their SRs and are classified as outliers of the \gstrp.
The lack of association for these sources 
could be due to a lower accuracy of the $\gamma$-ray position that might occur close to the Galactic plane
or to the systematic uncertainties of the diffuse emission model used in the 2FGL analysis.
The whole UGS sample will be reconsidered for associations with $\gamma$-ray blazar candidates
when the all-sky \wse\ survey will be available.

Assuming that all the 2FB blazar associations are correct, 
on the basis of our test (see Section~\ref{sec:test}),
we can argue that within our sample we would expect about 41 ($\sim$ 13.0\%) not recognized 
low-energy counterparts, 
for a total of 197 $\gamma$-ray blazar candidates within the 313 UGSs analyzed.

Finally, it is worth stressing that our association procedure provides also interesting information 
on the sources that do not have a $\gamma$-ray blazar candidates in the SR.
The absence of $\gamma$-ray blazar candidates selected according to our association procedure
could direct to better use the follow-up resources for identifying other $\gamma$-ray source candidates.
For example in the case of the unidentified $\gamma$-ray source: 2FGL J1446.8$-$4701
within the 1604 \wse\ sources that lie in its SR, we did not find any $\gamma$-ray blazar candidates.
This source has been recently identified with the pulsar 
PSR 1446-4701 (see Public List of LAT-Detected Gamma-Ray Pulsars)
\footnote{\underline{https://confluence.slac.stanford.edu/display/GLAMCOG/Public+List+of+LAT-Detected+Gamma-Ray+Pulsars}}.

\section{Comparison with other methods}
\label{sec:comparison}
We note that among the 313 UGSs analyzed, 
there are 70 sources that were also unidentified according to the investigation performed in the
first Fermi $\gamma$-ray catalog (1FGL), and 48 of them 
have been associated with a $\gamma$-ray blazar candidates in our analysis. 
In particular, a recent analysis of the 1FGL unidentified $\gamma$-ray sources has been carried out using two different statistical approaches: the Classification Tree
and the Logistic regression analyses \citep[see][and references therein]{ackermann11b}. 

For 44 out of the 48 UGSs, that have been analyzed on the basis 
of the above statistical methods, it is also possible to perform a comparison with our results
to verify if the 2FGL sources that we associated to a $\gamma$-ray blazar candidates have been also classified as AGNs
following the Ackermann et al. (2011b) procedures.
By comparing the results of our association method  with those in Ackermann et al. (2011b), we found that
27 out of 44 UGSs that we associate to a $\gamma$-ray blazar candidate are also classified as AGNs,
all of them with a probability higher than 71\% and 18 of them higher than 80\%. 
Among the remaining 17 out of 44 sources, 7 have been classified as pulsars,
with a very low probability with respect to the whole sample; in particular, 3 of these pulsar candidates 
are classified with a probability lower than 41\% and all of them lower than 71\%, making these classifications 
less reliable than those of the AGNs.
The last 10 UGSs did not have a classification in Ackermann et al. (2011b).
Consequently, we emphasize that our results are in good agreement with the classification 
suggested previously by Ackermann et al. (2011b)
consistent with the $\gamma$-ray blazar nature of the \wse\ candidates proposed in our analysis.

\section{Summary and Conclusions}
\label{sec:summary}
Recently, we discovered that blazars have peculiar mid-IR colors 
with respect to other galactic sources or different classes of AGNs.
In particular, we found that within the 3-dimensional IR parameter space they delineate a distinct, 
well-defined, region known as \wse\ Blazar Strip (Paper I).
Moreover, this distinction, mostly due to the non-thermal emission 
that dominates the IR radiation of blazars,
appears to be more evident when considering those blazars selected
on the basis of their $\gamma$-ray properties (Paper II)
so defining the \wse\ Gamma-ray blazar Strip (\gstrp).
Then, in Paper III, we built the \gstrp\ parametrization 
to test the consistency of the low energy counterpart of the AGUs, associated in 2FGL with the \gstrp.

On the basis of these results, in the present work, we developed a new association method to 
search for blazar counterparts of $\gamma$-ray sources 
and we applied this method to the blazars of the 2FGL sample. 
We also provide new $\gamma$-ray blazar candidates, potential counterparts of the UGSs, 
that lie within their $\gamma$-ray positional error region,
having the same mid-IR colors as the $\gamma$-ray blazars already associated.
We also tested our new procedure {\it a posteriori} trying to re-associate all the blazars in the 2FB sample and
we found that our results are in good agreement with different association procedures.

The application of our association procedure to the UGSs has led to 
the selection of possible blazar counterparts for 156 of 313 UGSs analyzed.

As also noted in Section~\ref{sec:ugs}, our association procedure provides also interesting information 
on the sources that do not have a $\gamma$-ray blazar candidates in the SRs as 
the case of the unidentified $\gamma$-ray source: 2FGL J1446.8$-$4701,
recently identified with the pulsar PSR 1446-4701.

Several developments will be considered to improve our association 
procedure, such as taking into account not only the IR colors,
correspondent to flux ratios, but also the IR fluxes as well as the IR-$\gamma$-ray spectral index correlation (Paper II)
and the sky distribution of the $\gamma$-ray blazar candidates, once the whole \wse\ data archive will be released. 
Then, it will be also possible to calibrate our association procedure choosing the different thresholds for the $s$
parameters at different Galactic latitudes to take into account of the \wse\ background. 

Moreover, our association method is complementary 
to those adopted in the 2FGL catalog analysis, because it is based on
different multifrequency information. For this reason, these methods could be 
in principle combined to increase the fraction of associated UGSs and the efficiency of the
association. Further developments of this new association 
method will be investigated in a forthcoming paper \citep{massaro12b}.

\acknowledgements
We thank the anonymous referee for the his/her comments.
F. Massaro is grateful S. Digel for their fruitful discussions for all the comments helpful toward improving our presentation. 
We also thank to A. Cavaliere, D. Harris, J. Grindlay,
J. Knodlseder, P. Giommi, N. Omodei, H.Smith and D. Thompson for their suggestions. 
The work at SAO and at Stanford University is supported in part by the NASA grant NNX10AD50G, NNH09ZDA001N and NNX10AD68G.
R. D'Abrusco gratefully acknowledges the financial 
support of the US Virtual Astronomical Observatory, which is sponsored by the
National Science Foundation and the National Aeronautics and Space Administration.
F. Massaro acknowledges the Fondazione Angelo Della Riccia for the grant awarded him to support 
his research at SAO during 2011 and the Foundation BLANCEFLOR Boncompagni-Ludovisi, n'ee Bildt
for the grant awarded him in 2010 to support his research.
TOPCAT\footnote{\underline{http://www.star.bris.ac.uk/$\sim$mbt/topcat/}} 
\citep{taylor2005} was used extensively in this work 
for the preparation and manipulation of the tabular data.
Part of this work is based on archival data, 
software or on-line services provided by the ASI Science Data Center.
This publication makes use of data products from the Wide-field Infrared Survey Explorer, 
which is a joint project of the University of California, Los Angeles, and the Jet Propulsion Laboratory/California Institute of Technology, 
funded by the National Aeronautics and Space Administration.

{}

\begin{table}
\tiny
\caption{UGS Associations.}
\begin{tabular}{|lccccccccccc|}
\hline
Name & distance (arcsec) & $c_{12}$ & $\sigma_{12}$ & $c_{23}$ & $\sigma_{23}$ & $c_{34}$ & $\sigma_{34}$ & $s_b$ & $s_q$ & $class$ & $\pi$ \\
\hline
\noalign{\smallskip}			         
2FGL J0038.8+6259      &        &      &      &      &      &      &      &      &      &    &   \\
\hline
  J003818.70+630605.2 & 443.08 & 1.15 & 0.03 & 2.59 & 0.03 & 2.52 & 0.03 & 0.89 & 0.99 & A & 2/830\\ 
  J003756.80+630459.2 & 492.33 & 0.67 & 0.04 & 2.09 & 0.06 & 2.36 & 0.12 & 0.40 & 0.00 & B & 3/830\\ 
  J003834.17+630621.7 & 413.68 & 0.95 & 0.06 & 2.55 & 0.09 & 2.69 & 0.16 & 0.22 & 0.19 & C & 0/830\\ 
\hline
2FGL J0158.6+8558      &        &      &      &      &      &      &      &      &      &    &   \\
\hline
  J014847.32+860345.3 & 707.33 & 1.21 & 0.03 & 3.10 & 0.03 & 2.66 & 0.05 & 0.29 & 0.73 & A & 1/2439\\
  J015619.63+855634.6 & 173.64 & 1.11 & 0.04 & 3.49 & 0.04 & 2.71 & 0.05 & 0.00 & 0.70 & B & 0/2439\\
\end{tabular}
\label{tab:main}
\end{table}

\begin{table}
\tiny\caption{UGS Associations.}
\begin{tabular}{|lccccccccccc|}
\hline
Name & distance (arcsec) & $c_{12}$ & $\sigma_{12}$ & $c_{23}$ & $\sigma_{23}$ & $c_{34}$ & $\sigma_{34}$ & $s_b$ & $s_q$ & $class$ & $\pi$ \\
\hline
\noalign{\smallskip}			         
2FGL J0416.0$-$4355      &        &      &      &      &      &      &      &      &      &    &   \\
\hline
  J041605.81$-$435514.6 & 47.18 & 1.13 & 0.04 & 2.94 & 0.04 & 2.37 & 0.08 & 0.58 & 0.58 & A & 0/2603\\ 
  J041442.89$-$434935.4 & 943.62 & 1.03 & 0.04 & 2.9 & 0.07 & 2.43 & 0.2 & 0.32 & 0.34 & B & 2/2603\\ 
  J041506.36$-$440958.9 & 1045.16 & 1.08 & 0.04 & 2.74 & 0.06 & 2.14 & 0.18 & 0.36 & 0.38 & B & 2/2603\\ 
  J041538.31$-$440412.8 & 568.43 & 1.01 & 0.05 & 2.6 & 0.1 & 2.49 & 0.3 & 0.21 & 0.23 & C & 1/2603\\ 
  J041709.78$-$435922.4 & 751.76 & 1.04 & 0.08 & 2.94 & 0.16 & 2.5 & 0.45 & 0.11 & 0.15 & C & 1/2603\\ 
\hline
2FGL J0420.9$-$3743      &        &      &      &      &      &      &      &      &      &    &   \\
\end{tabular}
\label{tab:main}
\end{table}

\begin{table}
\tiny\caption{UGS Associations.}
\begin{tabular}{|lccccccccccc|}
\hline
Name & distance (arcsec) & $c_{12}$ & $\sigma_{12}$ & $c_{23}$ & $\sigma_{23}$ & $c_{34}$ & $\sigma_{34}$ & $s_b$ & $s_q$ & $class$ & $\pi$ \\
\hline
\noalign{\smallskip}			         
2FGL J0608.3+2037      &        &      &      &      &      &      &      &      &      &    &   \\
\hline
  J060835.36+203604.1 & 237.43 & 1.03 & 0.03 & 2.83 & 0.03 & 2.22 & 0.05 & 0.76 & 0.77 & A & 0/1130\\ 
  J060831.76+203141.8 & 405.85 & 1.03 & 0.03 & 2.47 & 0.03 & 2.32 & 0.03 & 0.91 & 0.88 & A & 0/1130\\ 
  J060740.08+203959.2 & 579.55 & 1.11 & 0.04 & 2.86 & 0.04 & 2.37 & 0.09 & 0.57 & 0.58 & A & 0/1130\\ 
  J060803.38+204111.0 & 309.04 & 1.24 & 0.03 & 2.55 & 0.03 & 2.56 & 0.03 & 0.00 & 0.7 & B & 2/1130\\
  J060831.63+204122.5 & 260.5 & 1.19 & 0.04 & 2.82 & 0.09 & 2.74 & 0.12 & 0.18 & 0.32 & C & 0/1130\\
\hline
2FGL J0616.6+2425      &        &      &      &      &      &      &      &      &      &    &   \\
\end{tabular}
\label{tab:main}
\end{table}

\begin{table}
\tiny\caption{UGS Associations.}
\begin{tabular}{|lccccccccccc|}
\hline
Name & distance (arcsec) & $c_{12}$ & $\sigma_{12}$ & $c_{23}$ & $\sigma_{23}$ & $c_{34}$ & $\sigma_{34}$ & $s_b$ & $s_q$ & $class$ & $\pi$ \\
\hline
\noalign{\smallskip}			         
2FGL J0742.7$-$3113      &        &      &      &      &      &      &      &      &      &    &   \\
\hline
  J074226.42$-$305720.1 & 990.78 & 1.09 & 0.03 & 2.73 & 0.03 & 2.22 & 0.03 & 0.9 & 0.91 & A & 1/3499\\ 
  J074345.64$-$310758.0 & 873.05 & 1.02 & 0.05 & 2.29 & 0.05 & 2.16 & 0.15 & 0.31 & 0.25 & B & 1/3499\\
  J074303.55$-$312057.4 & 522.58 & 1.2 & 0.1 & 2.78 & 0.08 & 2.0 & 0.29 & 0.1 & 0.18 & C & 2/3499\\
\hline
2FGL J0744.1$-$2523      &        &      &      &      &      &      &      &      &      &    &   \\
\hline
  J074406.14$-$252154.1 & 142.98 & 1.27 & 0.03 & 2.55 & 0.03 & 1.98 & 0.05 & 0.00 & 0.44 & B & 0/630\\
  J074401.10$-$252203.9 & 180.53 & 1.26 & 0.06 & 2.77 & 0.04 & 2.66 & 0.08 & 0.00 & 0.5 & B & 0/630\\
\end{tabular}
\label{tab:main}
\end{table}

\begin{table}
\tiny\caption{UGS Associations.}
\begin{tabular}{|lccccccccccc|}
\hline
Name & distance (arcsec) & $c_{12}$ & $\sigma_{12}$ & $c_{23}$ & $\sigma_{23}$ & $c_{34}$ & $\sigma_{34}$ & $s_b$ & $s_q$ & $class$ & $\pi$ \\
\hline
\noalign{\smallskip}			         
2FGL J1243.9$-$6232      &        &      &      &      &      &      &      &      &      &    &   \\
\hline
  J124252.40$-$623214.0 & 437.79 & 1.13 & 0.04 & 2.52 & 0.04 & 2.52 & 0.04 & 0.63 & 0.64 & A & 0/854\\ 
  J124322.29$-$623912.7 & 466.0 & 0.7 & 0.08 & 2.41 & 0.07 & 2.29 & 0.15 & 0.3 & 0.09 & B & 0/854\\
  J124438.34$-$623425.2 & 317.58 & 0.85 & 0.06 & 2.3 & 0.11 & 2.6 & 0.15 & 0.24 & 0.14 & C & 0/854\\
\hline
2FGL J1248.6$-$5510      &        &      &      &      &      &      &      &      &      &    &   \\
\hline
  J124946.06$-$550758.3 & 602.9 & 1.03 & 0.05 & 2.7 & 0.07 & 2.45 & 0.15 & 0.36 & 0.36 & B & 2/4244\\ 
\hline
2FGL J1255.8$-$5828      &        &      &      &      &      &      &      &      &      &    &   \\
\end{tabular}
\label{tab:main}
\end{table}

\begin{table}
\tiny\caption{UGS Associations.}
\begin{tabular}{|lccccccccccc|}
\hline
Name & distance (arcsec) & $c_{12}$ & $\sigma_{12}$ & $c_{23}$ & $\sigma_{23}$ & $c_{34}$ & $\sigma_{34}$ & $s_b$ & $s_q$ & $class$ & $\pi$ \\
\hline
\noalign{\smallskip}			         
2FGL J1517.2+3645      &        &      &      &      &      &      &      &      &      &    &   \\
\hline
  J151649.27+365022.9 & 420.93 & 1.05 & 0.05 & 2.63 & 0.09 & 2.48 & 0.24 & 0.27 & 0.27 & B & 0/1352\\ 
  J151752.13+364125.5 & 508.82 & 1.04 & 0.04 & 3.01 & 0.06 & 2.19 & 0.19 & 0.34 & 0.36 & B & 0/1352\\ 
\hline
2FGL J1518.4$-$5233      &        &      &      &      &      &      &      &      &      &    &   \\
\hline
  J151807.33$-$523431.0 & 184.75 & 0.66 & 0.07 & 2.31 & 0.05 & 1.41 & 0.17 & 0.26 & 0.13 & B & 0/979\\
\hline
2FGL J1543.7$-$0241      &        &      &      &      &      &      &      &      &      &    &   \\
\hline
  J154315.19$-$022531.4 & 1052.36 & 1.06 & 0.08 & 2.85 & 0.18 & 2.58 & 0.43 & 0.11 & 0.14 & C & 0/3462\\
\end{tabular}
\label{tab:main}
\end{table}

\begin{table}
\tiny\caption{UGS Associations.}
\begin{tabular}{|lccccccccccc|}
\hline
Name & distance (arcsec) & $c_{12}$ & $\sigma_{12}$ & $c_{23}$ & $\sigma_{23}$ & $c_{34}$ & $\sigma_{34}$ & $s_b$ & $s_q$ & $class$ & $\pi$ \\
\hline
\noalign{\smallskip}			         
2FGL J1623.2+4328      &        &      &      &      &      &      &      &      &      &    &   \\
\hline
  J162324.12+432533.8 & 183.07 & 0.91 & 0.04 & 2.64 & 0.06 & 2.47 & 0.14 & 0.41 & 0.35 & B & 0/1983\\ 
  J162237.62+433801.8 & 720.08 & 1.13 & 0.05 & 2.92 & 0.08 & 2.25 & 0.24 & 0.29 & 0.3 & B & 0/1983\\ 
\hline
2FGL J1624.2$-$2124      &        &      &      &      &      &      &      &      &      &    &   \\
\hline
  J162608.57$-$211709.1 & 1617.78 & 1.38 & 0.04 & 2.79 & 0.06 & 2.57 & 0.12 & 0.00 & 0.4 & B & 0/17861\\
\hline
2FGL J1627.8+3219      &        &      &      &      &      &      &      &      &      &    &   \\
\hline
  J162800.41+322414.9 & 319.07 & 1.12 & 0.05 & 2.89 & 0.09 & 2.52 & 0.27 & 0.21 & 0.26 & C & 1/1045\\
\end{tabular}
\label{tab:main}
\end{table}

\begin{table}
\tiny\caption{UGS Associations.}
\begin{tabular}{|lccccccccccc|}
\hline
Name & distance (arcsec) & $c_{12}$ & $\sigma_{12}$ & $c_{23}$ & $\sigma_{23}$ & $c_{34}$ & $\sigma_{34}$ & $s_b$ & $s_q$ & $class$ & $\pi$ \\
\hline
\noalign{\smallskip}			         
2FGL J1726.6$-$3545      &        &      &      &      &      &      &      &      &      &    &   \\
\hline
  J172706.54$-$354400.0 & 332.55 & 0.83 & 0.05 & 2.26 & 0.05 & 1.95 & 0.16 & 0.38 & 0.16 & B & 6/2747\\
  J172645.15$-$355126.8 & 340.31 & 0.78 & 0.04 & 1.97 & 0.04 & 2.01 & 0.06 & 0.52 & 0.00 & B & 6/2747\\
  J172743.40$-$355104.9 & 823.97 & 0.75 & 0.05 & 1.97 & 0.06 & 1.8 & 0.14 & 0.38 & 0.16 & B & 6/2747\\
\hline
2FGL J1727.6+0647      &        &      &      &      &      &      &      &      &      &    &   \\
\hline
  J172644.95+063918.6 & 964.35 & 1.1 & 0.05 & 2.91 & 0.08 & 2.46 & 0.21 & 0.28 & 0.3 & B & 0/3011\\ 
  J172743.08+063729.0 & 610.69 & 0.88 & 0.06 & 2.79 & 0.13 & 2.16 & 0.41 & 0.15 & 0.16 & C & 0/3011\\
\end{tabular}
\label{tab:main}
\end{table}

\begin{table}
\tiny\caption{UGS Associations.}
\begin{tabular}{|lccccccccccc|}
\hline
Name & distance (arcsec) & $c_{12}$ & $\sigma_{12}$ & $c_{23}$ & $\sigma_{23}$ & $c_{34}$ & $\sigma_{34}$ & $s_b$ & $s_q$ & $class$ & $\pi$ \\
\hline
\noalign{\smallskip}			         
2FGL J1742.5$-$3323      &        &      &      &      &      &      &      &      &      &    &   \\
\hline
  J174220.74$-$333005.4 & 414.44 & 0.55 & 0.11 & 2.26 & 0.08 & 2.14 & 0.06 & 0.28 & 0.07 & B & 13/2475\\
\hline
2FGL J1743.2$-$2304      &        &      &      &      &      &      &      &      &      &    &   \\
\hline
  J174224.66$-$225942.4 & 710.92 & 1.02 & 0.06 & 2.56 & 0.04 & 1.88 & 0.04 & 0.58 & 0.35 & B & 0/3656\\
\hline
2FGL J1745.6+0203      &        &      &      &      &      &      &      &      &      &    &   \\
\hline
  J174526.95+020532.7 & 208.91 & 1.06 & 0.04 & 2.63 & 0.04 & 2.19 & 0.08 & 0.57 & 0.57 & A & 0/4745\\ 
  J174647.05+020925.9 & 1066.59 & 1.07 & 0.04 & 2.61 & 0.06 & 2.27 & 0.16 & 0.38 & 0.39 & A & 0/4745\\
\end{tabular}
\label{tab:main}
\end{table}

\begin{table}
\tiny\caption{UGS Associations.}
\begin{tabular}{|lccccccccccc|}
\hline
Name & distance (arcsec) & $c_{12}$ & $\sigma_{12}$ & $c_{23}$ & $\sigma_{23}$ & $c_{34}$ & $\sigma_{34}$ & $s_b$ & $s_q$ & $class$ & $\pi$ \\
\hline
\noalign{\smallskip}			         
2FGL J1802.8$-$6706      &        &      &      &      &      &      &      &      &      &    &   \\
\hline
  J180100.62$-$670503.7 & 647.12 & 1.0 & 0.04 & 3.12 & 0.06 & 2.47 & 0.13 & 0.00 & 0.41 & B & 1/3111\\
\hline
2FGL J1811.3$-$2421      &        &      &      &      &      &      &      &      &      &    &   \\
\hline
  J181126.39$-$240459.9 & 1018.47 & 0.84 & 0.04 & 2.74 & 0.03 & 2.56 & 0.03 & 0.65 & 0.56 & A & 0/4178\\
\hline
2FGL J1813.6$-$2821      &        &      &      &      &      &      &      &      &      &    &   \\
\hline
  J181345.24$-$283058.6 & 594.38 & 0.75 & 0.04 & 2.26 & 0.03 & 2.1 & 0.02 & 0.94 & 0.4 & A & 0/1731\\
\hline
2FGL J1819.3$-$1523      &        &      &      &      &      &      &      &      &      &    &   \\
\end{tabular}
\label{tab:main}
\end{table}

\begin{table}
\tiny\caption{UGS Associations.}
\begin{tabular}{|lccccccccccc|}
\hline
Name & distance (arcsec) & $c_{12}$ & $\sigma_{12}$ & $c_{23}$ & $\sigma_{23}$ & $c_{34}$ & $\sigma_{34}$ & $s_b$ & $s_q$ & $class$ & $\pi$ \\
\hline
\noalign{\smallskip}			         
2FGL J1835.4+1349      &        &      &      &      &      &      &      &      &      &    &   \\
\hline
  J183522.00+135733.9 & 484.14 & 1.15 & 0.06 & 2.78 & 0.06 & 2.28 & 0.13 & 0.33 & 0.38 & A & 1/2344\\ 
  J183535.35+134848.9 & 141.16 & 0.78 & 0.04 & 2.26 & 0.05 & 2.02 & 0.15 & 0.44 & 0.19 & B & 1/2344\\
  J183539.22+135055.0 & 206.92 & 1.22 & 0.04 & 2.82 & 0.06 & 2.67 & 0.11 & 0.19 & 0.44 & B & 1/2344\\
\hline
2FGL J1835.5$-$0649      &        &      &      &      &      &      &      &      &      &    &   \\
\hline
  J183538.53$-$064854.2 & 71.66 & 0.61 & 0.04 & 1.74 & 0.04 & 2.12 & 0.09 & 0.55 & 0.00 & B & 0/622\\
  J183554.30$-$065518.0 & 427.7 & 0.74 & 0.05 & 1.88 & 0.05 & 1.96 & 0.09 & 0.39 & 0.00 & B & 0/622\\
\end{tabular}
\label{tab:main}
\end{table}

\begin{table}
\tiny\caption{UGS Associations.}
\begin{tabular}{|lccccccccccc|}
\hline
Name & distance (arcsec) & $c_{12}$ & $\sigma_{12}$ & $c_{23}$ & $\sigma_{23}$ & $c_{34}$ & $\sigma_{34}$ & $s_b$ & $s_q$ & $class$ & $\pi$ \\
\hline
\noalign{\smallskip}			         
2FGL J1917.0$-$3027      &        &      &      &      &      &      &      &      &      &    &   \\
\hline
  J191637.70$-$303356.6 & 525.92 & 1.19 & 0.06 & 3.01 & 0.08 & 2.47 & 0.2 & 0.15 & 0.28 & C & 1/1040\\ 
\hline
2FGL J1923.4+2013      &        &      &      &      &      &      &      &      &      &    &   \\
\hline
  J192142.39+201107.1 & 1491.67 & 1.08 & 0.12 & 2.61 & 0.07 & 2.29 & 0.13 & 0.28 & 0.27 & B & 7/14898\\ 
  J192540.72+201244.2 & 1870.96 & 0.95 & 0.05 & 2.64 & 0.04 & 1.88 & 0.06 & 0.55 & 0.34 & B & 7/14898\\
  J192501.65+204022.6 & 2080.26 & 0.95 & 0.06 & 2.28 & 0.04 & 2.01 & 0.08 & 0.43 & 0.22 & B & 7/14898\\
\hline
2FGL J1924.9$-$1036      &        &      &      &      &      &      &      &      &      &    &   \\
\end{tabular}
\label{tab:main}
\end{table}

\begin{table}
\tiny\caption{UGS Associations.}
\begin{tabular}{|lccccccccccc|}
\hline
Name & distance (arcsec) & $c_{12}$ & $\sigma_{12}$ & $c_{23}$ & $\sigma_{23}$ & $c_{34}$ & $\sigma_{34}$ & $s_b$ & $s_q$ & $class$ & $\pi$ \\
\hline
\noalign{\smallskip}			         
2FGL J2009.2$-$1505      &        &      &      &      &      &      &      &      &      &    &   \\
\hline
  J200901.29$-$151620.3 & 700.87 & 1.02 & 0.06 & 3.02 & 0.1 & 2.52 & 0.24 & 0.17 & 0.26 & C & 0/1969\\ 
\hline
2FGL J2031.4$-$1842      &        &      &      &      &      &      &      &      &      &    &   \\
\hline
  J203142.40$-$182208.0 & 1219.37 & 0.89 & 0.04 & 2.39 & 0.09 & 2.31 & 0.26 & 0.29 & 0.18 & B & 0/3790\\
  J203159.63$-$183824.4 & 499.26 & 0.92 & 0.07 & 2.66 & 0.11 & 2.3 & 0.32 & 0.22 & 0.18 & C & 1/3790\\ 
  J203003.91$-$184441.6 & 1208.58 & 1.23 & 0.06 & 2.85 & 0.12 & 2.43 & 0.32 & 0.11 & 0.23 & C & 1/3790\\
  J203035.02$-$185910.1 & 1269.41 & 1.24 & 0.06 & 2.95 & 0.1 & 2.63 & 0.2 & 0.12 & 0.27 & C & 1/3790\\
\hline
2FGL J2044.4$-$4757      &        &      &      &      &      &      &      &      &      &    &   \\
\hline
  J204444.65$-$481006.7 & 764.03 & 1.02 & 0.03 & 2.8 & 0.04 & 2.44 & 0.07 & 0.64 & 0.65 & A & 0/1279\\ 
  J204520.57$-$475648.5 & 515.06 & 1.39 & 0.04 & 2.89 & 0.06 & 2.41 & 0.14 & 0.00 & 0.38 & B & 0/1279\\
\hline
2FGL J2131.0$-$5417      &        &      &      &      &      &      &      &      &      &    &   \\
\hline
  J213014.80$-$540748.6 & 728.48 & 1.09 & 0.05 & 2.6 & 0.13 & 2.64 & 0.32 & 0.16 & 0.18 & C & 1/2080\\ 
\hline
2FGL J2200.1$-$6931      &        &      &      &      &      &      &      &      &      &    &   \\
\hline
  J215916.12$-$692856.1 & 324.01 & 1.11 & 0.05 & 2.8 & 0.09 & 2.3 & 0.28 & 0.25 & 0.26 & B & 0/910\\ 
\hline
\noalign{\smallskip}			           
\end{tabular}
\label{tab:main}
\end{table}

\end{document}